\documentstyle{article}
\begin{document}                                                                                   

\centerline{\bf ON CALCULATION OF MICROLENSING LIGHT CURVE}
\centerline{\bf BY GRAVITATIONAL LENS CAUSTIC}

\vspace{0.5cm}

\centerline{\bf Mikhail B. Bogdanov}

\vspace{0.5cm}

\centerline{Saratov State University, Bolshaya Kazachiya st., 112-a,}
\centerline{410026 Saratov, Russia}

\vspace{0.5cm}

\noindent {\bf Abstract.} For an analysis of microlensing observational data in
case of 
binary gravitational lenses as well as for an interpretation of observations
of high magnification events in multiple images of a lensed quasar it is
necessary to calculate for a given source the microlensing light curve by a
fold caustic. This problem comes to the numerical calculation of a singular
integral. We formulated the sufficient condition of a convergence of the
integral sum for this singular integral. The strictly approach to the problem
of a comparison of model results with the unequally sampled observational
data consists in calculation of the model light curve in equidistant points of
the canonical dissection of the integration segment and a following
interpolation of its values at the moments of observations.

\vspace{1cm}

The high angular resolution observations in optical and infrared wave bands
are of great importance for the solution of many problems of astronomy and 
astrophysics: the search for close binaries, the measurement of angular 
diameters of stars, the investigation of brightness distributions across the
stellar disks, the study of a fine spatial structure of active galactic
nuclei (AGN). A new tool to achieving of a high angular resolution has been
realized recently -- the observations of gravitational microlensing. 

In some microlensing events observed by MACHO, OGLE, and PLANET 
collaborations perhaps there are the binary gravitational lenses. For these
lenses the character of the gravitational potential creates caustics in the
source plane. The crossing of a caustic leads to sharp variation of a flux
received from the lensed source. The character of this variation is depend on
the source angular structure. Thus the observed microlensing light curve
contains an information about the source brightness distribution (Albrow 
{\it et al.}, 1999; Gaudi and Gould, 1999; Bogdanov and Cherepashchuk, 2000). 

The high magnification events in fluxes received from multiple images of a 
lensed quasar due to gravitational microlensing by stars in a lens galaxy are 
connected also with the complex caustic system which is created by the total 
gravitational potential (Schneider {\it et al.}., 1992; Zakharov, 1997). The 
observations of these events allow to obtain an information about the quasar 
central engine -- the accretion disk, which surrounds a super massive black
hole in the galactic nucleus (Grieger {\it et al.}, 1991; Mineshige and
Yonehara, 1999; Agol and Krolik, 1999; Bogdanov and Cherepashchuk, 2001). The
possible angular resolution that can be realized in the caustic crossing
microlensing observations is measured by the value of order or less of
microsecond of arc.

From the general considerations we can believe that the case of microlensing
by a fold caustic is a more probable (Gaudi and Gould, 1999). The angular
sizes of observed stars of the bulge of Galaxy or the Magellanic Clouds as
well as the sizes of AGN's accretion disks are very small. Therefore we may
neglect of a caustic curvature and regard the fold caustic as a
straightforward line. In this case the picture of microlensing depends only
on one-dimensional strip brightness distribution $B(x)$ in the direction of
$x$ axis that is perpendicular to caustic. It is known that $B(x)$ is
connected with a two-dimensional brightness distribution $b(x,y)$ by the
integral equation

$$
B(x)=\int\limits_{-\infty}^{\infty} \int\limits_{-\infty}^{\infty}
b(\xi ,y)\delta(\xi -x)\,d\xi dy , \eqno (1)
$$

\noindent where $\delta (x)$ is the Dirac function. For a circularly 
symmetric source
the brightness depends only on the radial distance to source center $r$. In
this case $B(x)$ is connected with $b(r)$ by the Abel's integral equation

$$
B(x)=\int\limits_{x}^{\infty} \frac {2b(r)r\,dr} {\sqrt{r^2-x^2}} .\eqno (2)
$$

\noindent For real astronomical objects the strip brightness distribution is a 
non-negative, continuous, and smooth enough function.

If $\xi$ is the angular distance of point with the brightness $B(\xi)$ to 
center of source and $x$ is the angular distance of source center to caustic 
then the coefficient of amplification can be written as (Schneider {\it et 
al.},1992; Zakharov, 1997):

$$
A(x-\xi)=A_{0}+\frac {K} {\sqrt{x-\xi}} H(x-\xi) , \eqno (3)
$$
    
\noindent where $H(x-\xi)$ is the Heaviside step function ($H = 0$ for
negative and 
$H = 1$ for non-negative values of its argument), $A_0$ and $K$ are certain 
constants for the given caustic crossing. The observed microlensing light
curve $I(x)$ as a function of distance to caustic $x$ can be expressed by the 
convolution integral equation:

$$
I(x)=A(x)*B(x)=\int\limits_{-\infty}^\infty A(x-\xi )B(\xi )\,d\xi  . 
\eqno (4)
$$

We can admit further without a lack of community that in the expression (3) 
$A_0 = 0$ and $K = 1$. Far from a caustic the integrand in the expression (4) 
is small enough and we can consider its to be equal to zero for $\xi < x_1$ . 
Then as it follows from equations (3) and (4) the estimation of flux value for
a given strip brightness distribution when $x = x_n$ comes to the numerical 
calculation of the singular integral

$$
I(x_n)=\int\limits_{x_1}^{x_n}\frac {B(\xi)\,d\xi } {\sqrt{x_n-\xi }}.\eqno (5)
$$

It is known that the numerical estimation of a singular integral requires of 
certain precautions. The similar problems were examined by Belotzerkovski and 
Lifanov (1985). In particular it can be shown that an attempt to calculate 
$I(x_i)$ in arbitrary spaced points $x_i$ when $B(\xi_i)$ is given in 
equidistant points $\xi_i$ leads to a large error of the result. We formulate 
below the sufficient condition of convergence of the integral sum for the
singular integral of our type.

At first we remind the basic definitions and one important theorem.

\vspace{0.5cm}

{\it Definition 1}

\vspace{0.5cm}

Let the segment $[x_1,x_n]$ is divided by points $x_i, i = 1,2,\ldots n$ into 
$n-1$ equal sub-segments of length $h$, and points $\xi_i , i = 1,2,\ldots n-
1$ are the middle points of these sub-segments: $\xi_i = (x_i + x_{i+1})/2$ . 
Then they say that the sequences of points $x_i$ and $\xi_i$ create the 
canonical dissection of this segment.

\vspace{0.5cm}

{\it Definition 2}

\vspace{0.5cm}

Let a function $f(x)$ is determined in a set $D$, and for two arbitrary
values of its argument $x_1,x_2\in D$ is valid the inequality

$$
                |f(x_1)-f(x_2)| \le k |x_1-x_2|^\alpha ,
$$

\noindent where $k$ and $\alpha $ are positive numbers, and $0 < \alpha \le 1$. 
Then they say that $f(x)$ satisfies the H\"older's condition of power 
$\alpha $ with the coefficient $k$ in the set $D$.

\vspace{0.5cm}

{\it Theorem 1 (The mean value theorem)}

\vspace{0.5cm}

Let functions $f(x)$ and $g(x)$ are both integrable in a segment $[a,b]$, and 
$f(x)$ is bounded $m \le f(x) \le M$, and sign of $g(x)$ is invariable in
this segment. Then the product of these functions $f(x)g(x)$ is also
integrable, and

$$
\int\limits_{a}^{b} f(x)g(x)\,dx=\mu\int\limits_{a}^{b} g(x)\,dx , 
$$

\noindent where $m \le \mu \le M$ .

\vspace{0.5cm}

\noindent The proof of this theorem can be found in textbooks on mathematical
analysis.

Take into consideration these definitions and the mean value theorem we can 
formulated the sufficient condition of convergence of integral sum for the 
rectangle rule that correspond to the singular integral (5) as the following
{\it Theorem.}

\vspace{0.5cm}

{\it Theorem}

\vspace{0.5cm}

Let a function $B(x)$ satisfies the H\"older's condition of power $\alpha $  
in segment $[x_1,x_n]$, and sequences of points $x_i , i = 1,2,\ldots n$ and 
$\xi_i , i = 1,2,\ldots n-1$ create the canonical dissection of this segment 
with the step $h$. Then

$$
\Delta =\left|\int\limits_{x_1}^{x_n}\frac {B(\xi)\,d\xi } {\sqrt{x_n-\xi }}-
\sum_{i=1}^{n-1} \frac {B(\xi_i)h} {\sqrt{x_n-\xi_i }} \right| \le 
o(h^{\alpha+1/2}) . 
$$

\vspace{0.5cm}

{\it Proof}

\vspace{0.5cm}

The sequence of points $x_i$ divides according to {\it Definition 1} the 
segment $[x_1,x_n]$ into $n-1$ sub-segments of length $h$. In each of these 
sub-segments the function $B(\xi)$ is continuous and bounded, and the sign of 
function $g(\xi) = (x_n - \xi )^{-1/2}$ is invariable. Using {\it Theorem 1} 
and integrate of the function $g(\xi )$ in each sub-segment we can to write

$$
\Delta =\left|\sum_{i=1}^{n-1} 2B(\xi_i^*)\left[\sqrt{x_n-x_i }-\sqrt{x_n-
x_{i+1}}\,\right]-\sum_{i=1}^{n-1} \frac {B(\xi_i)h} {\sqrt{x_n-\xi_i }}\right| , 
$$
   
\noindent where $\xi_i^*$ is a certain point in sub-segment of number $i$ .
Further we have

$$
\Delta =\left|\sum_{i=1}^{n-1} \frac {2B(\xi_i^*)\sqrt{x_n-\xi_i} \left[ 
\sqrt{x_n-x_i }-\sqrt{x_n-x_{i+1}}\,\right] - B(\xi_i)h} {\sqrt{x_n-\xi_i }} 
\right| . 
$$
   
Let we multiply and divide simultaneously the expression that closed in
brackets by factor $\left( \sqrt{x_n-x_i } + \sqrt{x_n-x_{i+1}}\,\right)$.
Then we have

$$
\left[ \sqrt{x_n-x_i }-\sqrt{x_n-x_{i+1}}\,\right] = \frac {x_{i+1}-x_i} { 
\sqrt{x_n-x_i } + \sqrt{x_n-x_{i+1}}} \le \sqrt{h} .
$$
    
\noindent The last inequality is valid so far as $x_{i+1} - x_i = h$, and the
denominator of this fraction is minimal when $i = n - 1$. 

The common denominator in the expression for $\Delta $ is also minimal when 
$i = n - 1$, and its minimal value is equal to $\sqrt{h/2}$ according to
{\it Definition 1}. Thus we have restriction

$$
\Delta \le \left|\sum_{i=1}^{n-1} \frac {2B(\xi_i^*)\sqrt {h/2} \sqrt{h} - 
B(\xi_i)h} {\sqrt{h/2} }\right| \le \sum_{i=1}^{n-1} 
\sqrt{2h}\left|\sqrt{2}B(\xi_i^*) - B(\xi_i) \right| . 
$$

\noindent Take into account {\it Definition 2} we have finally
$\Delta \le o(h^{\alpha + 1/2})$. The {\it Theorem} is proved.

\vspace{0.5cm}

{\it Remark 1}

\vspace{0.5cm}

The H\"older's condition is a more strong restriction on the behavior of a 
function in comparison with a continuity. If the H\"older's condition is
valid then the function is continuous. The opposite statement is generally
wrong. It is obvious that for implementation of the H\"older's condition
when $\alpha = 1$ sufficiently to demand of bounded first differential of
the function. This demand is fulfil for strip brightness distributions $B(x)$
of real astronomical objects. 

\vspace{0.5cm}

{\it Remark 2}

\vspace{0.5cm}

For increase of precision of the integral sum calculation in the same
sequence of points $x_i$ the number of points $\xi_i$ can be multiplied by
factor $k > 1$, where $k$ is entire. The new sequence $\tilde \xi_i$ can
determined as $\tilde \xi_i  = x_1 + h(i-1/2)/k , i = 1,2,\ldots k(n-1)$ .
This case is equivalent to the estimation of the singular integral in a new
sequence of points $\tilde x_i = x_1 + i h/k , i = 1,2,\ldots k(n-1)$. The
new points $\tilde x_i$ are partly coincide with points of the previous
sequence $x_i$ and partly located between them. It is clear that the new
sequences $\tilde x_i$ and $\tilde \xi_i $ also create the canonical
dissection of segment $[x_1,x_n]$ .

\vspace{0.5cm}

{\it Remark 3}

\vspace{0.5cm}

The observed values of the flux in microlensing observations usually have 
unequal sampling intervals. Therefore the strictly approach to the problem of 
comparison of model results with the observational data consists in the 
calculation of values of the model microlensing light curve in equidistant
points of the canonical dissection of the integration segment and following
interpolation of them at the moments of observations.

\vspace{0.5cm}

Author is grateful to A.F.Zakharov for the attraction of his attention to
this problem.

\vspace{1.5cm}

\centerline{\bf References}

\vspace{0.5cm}

\noindent Agol E., Krolik J.: 1999, Astrophys. J. {\bf 524}, 49.

\noindent Albrow M.D., Beaulieu J.-P., Caldwell J.A.R. et al.: 1999,
Astrophys. J. {\bf 512}, 672.

\noindent Belotzerkovski S.M., Lifanov I.K.: 1985, Numerical methods in
singular integral equations. Nauka, Moscow (in Russian).

\noindent Bogdanov M.B., Cherepashchuk A.M.: 2000, Astron. Rept. {\bf 44},
745. 

\noindent Bogdanov M.B., Cherepashchuk A.M.: 2001, Astrophys. and Space Sci.,
in press. 

\noindent Gaudi S.B., Gould A.: 1999, Astrophys. J. {\bf 513}, 619.

\noindent Grieger B., Kayser R., Schramm T.: 1991, Astron. and Astrophys.
{\bf 252}, 508.

\noindent Mineshige S., Yonehara A.: 1999, Publ. Astron. Soc. Japan.
{\bf 51}, 497.

\noindent Schneider P., Ehlers J., Falco E.E.: 1992, Gravitational lenses.
Springer, Berlin.

\noindent Zakharov A.F.: 1997, Gravitational lenses and microlenses.
Yanus-K, Moscow (in Russian)

\end{document}